# Data Vision:
# Learning to See Through Algorithmic Abstraction

**Samir Passi**
Dept. of Information Science
Cornell University
sp966@cornell.edu

**Steven J. Jackson**
Dept. of Information Science
Cornell University
sjj54@cornell.edu

**ABSTRACT**

Learning to see through data is central to contemporary forms of algorithmic knowledge production. While often represented as a mechanical application of rules, making algorithms work with data requires a great deal of situated work. This paper examines how the often-divergent demands of mechanization and discretion manifest in data analytic learning environments. Drawing on research in CSCW and the social sciences, and ethnographic fieldwork in two data learning environments, we show how an algorithm's application is seen sometimes as a mechanical sequence of rules and at other times as an array of situated decisions. Casting data analytics as a rule-*based* (rather than rule-*bound*) practice, we show that effective data vision requires would-be analysts to straddle the competing demands of formal abstraction and empirical contingency. We conclude by discussing how the notion of data vision can help better leverage the role of human work in data analytic learning, research, and practice.

**ACM Classification Keywords**
H.m. [Information Systems]: Miscellaneous

**INTRODUCTION**

Algorithmic data analysis has come to enable new ways of producing and validating knowledge [15, 25]. Algorithms are integral to many contemporary knowledge practices, especially ones that rely on the analysis of large-scale datasets [15, 20, 21, 34]. At the same time, we know that algorithms can be selective [34], subjective [7], and biased [3]; that they work on multiple assumptions about the world and how it functions [5, 15, 34, 35]; and that they simultaneously enable and constrain possibilities of human action and knowledge [5, 6]. Algorithmic knowledge production is a deeply social and collaborative practice with sociocultural, economic, and political groundings and consequences.

In all these ways, data analysis embodies a distinct and powerful way of *seeing* the world. Data analysts learn to represent and organize the world through computational forms such as graphs, matrices, and a host of standardized formats, enabling them to make knowledge claims based on algorithmic analyses. But this is just one half of the story. The world doesn't always neatly fit into spreadsheets, matrices, and tables. While data analysis is often understood as the work of faceless and unbiased numbers and algorithms, a large amount of situated and discretionary work is required to organize and manipulate the world algorithmically. Effective algorithmic analysis also demands mastery of the ways that worlds and tools are put together, and *which* worlds and tools are so combined (across the wide range of methods, tools, and objects amenable to representation). Taken together, these two seemingly contradictory features constitute what we call *data vision:* the ability to organize and manipulate the world with data and algorithms, while simultaneously mastering forms of discretion around *why*, *how*, and *when* to apply and improvise around established methods and tools in the wake of empirical diversity.

Integrated, often seamlessly, in the practice of expert practitioners, these contradictory demands stand out with particular clarity in the moments of learning and professionalization through which novices learn to master and balance the intricacies of data vision. How do students learn to "see" the world through data and algorithms? How do they learn to maneuver and improvise around forms and formalizations in the face of empirical contingency? This paper addresses such questions in the context of data analytic learning environments such as classrooms and workshops.

While distinct from other contexts of professional practice (e.g., industry settings or research centers), learning environments provide partial but meaningful sites to understand some of the ways in which would-be practitioners are immersed and acculturated into professional discourse and practice. [On the relation and relevance of learning environments for 'mature' professional practice, see inter alia 8, 16, and 24]. The explicit focus in learning environments on demonstrating established methods and theories to would-be professionals allows us to see how particular pedagogic demonstrations and analytic examples enable specific algorithmic norms and heuristics. More importantly, a study of classrooms and workshops draws attention to the social aspects of learning – a process of participation and membership in a discourse, instead of just a set of individual experiences. In learning environments, aspects of professionalization are accomplished through guided interactions between instructors, students, teaching assistants, educational materials, assignments, and exams. Learning environments thus function as important sites in which would-be data analysts learn to see the world through and as data – a crucial rite of passage on their way to becoming full-fledged members in the data analytic "community of practice." [24]

This paper describes two separate sequences of events – one from a machine-learning classroom, and another from a series of digital humanities workshops – to show how learning to see through data requires students to maintain a balance between viewing the world through abstract constructs, while simultaneously adapting to empirical contingency. We advance a rule-*based* (as opposed to a rule-*bound*) understanding of data analytic practice, highlighting the situated interplay between formal abstraction and mechanical routinization on the one hand, and discretionary action and empirical contingency on the other. We show how it is the mastery of this interplay – and not just the practice of data analytic techniques in their formal dimension – that is central to the growing skill and efficacy of would-be data analysts. We argue that better understanding of data vision in its more comprehensive and discretionary forms can help researchers and instructors better engage and leverage the human dimensions and limits of data analytic learning and practice.

The sections that follow begin by reviewing CSCW, HCI, and social science literatures on professional vision, situated knowledge, and discretionary practice. We then describe our research sites, before moving to the empirical examples. We conclude by discussing the implications of the notion and practice of data vision, and the distinction between a rule-bound and rule-based understanding of data analysis, for data analytic learning and practice, and for CSCW research and practice more broadly.

**PROFESSIONAL VISION, SITUATED KNOWLEDGE, AND DISCRETIONARY PRACTICE**

Our work on data vision builds on a classic and growing body of work in the social sciences that has explored forms of identity, practice, and perception underpinning and constituting forms of professional knowledge. Goodwin's work on *professional vision* [16] analyzes two professional activities (archaeological field excavation and legal argumentation) to show how professionals learn to "see" relevant objects of professional knowledge with and through *practice*: the exposure to and exercise of theories, methods, and tools to produce artifacts and knowledge in line with professional goals. Learning professional practice, he argues, help professionals make salient specific aspects of phenomena, transforming them into objects of knowledge amenable to professional analysis. Learning to see the world professionally, however, is not reducible to the mastery of generic rules and formal techniques. Instead, professional vision is slowly and carefully built through training, socialization, and immersion into professional discourse [16, 24, 30, 32]. Professional vision, thus, is a substantive and collaborative sociocultural accomplishment – a way of seeing the world constructed and shaped by a "community of practice." [24]

A key aspect of professional vision, as Abbott [1] argues, is the way in which practitioners situate given problems within existing repertoires of professional knowledge, methods, and expertise. According to Abbott, the process of situating given problems – of "seeing" professionally – must be clear enough for professionals to create relations between a given problem and existing knowledge *(e.g., what can I say about this specific dataset?)*, yet abstract, even ambiguous, enough to enable professionals to create such relations for a wide variety of problems *(e.g., what are the different kinds of datasets about which I can say something?)*.

A similar interplay between abstraction, clarity, and discretion exists within data analytic practices. Algorithms, developed in computational domains such as machine-learning, information retrieval, and natural language processing, provide means of analyzing data. It is often argued that a specific algorithm can work on multiple datasets as long as the datasets are modeled in particular ways. However, data analysis requires much more work than simply applying an algorithm to a dataset. As Mackenzie argues: certain data analytic practices such as vectorization, approximation, and modeling often mask the inherent subjectivity of dataset and algorithms, imbuing them with a sense of inherent "generalization." [26] From choice of analytic method, to choices concerning data formatting, to decisions about how best to represent and communicate data analytic results to 'outside' audiences, a large amount of situated and discretionary work – e.g., in the form of data collection, data cleaning, data modeling, and other forms of pre- and post-processing – is required to make datasets work with chosen algorithms. Data analysts not just learn to see and organize the world through data and algorithms, but also learn and discern meaningful and effective combinations of data and algorithms. As Gitelman et al. [15] argue: "raw data" – at least as a workable entity – is an oxymoron. It takes *work* to make data work.

Abbott's [1] example of chess is instructive in evoking the situated and discretionary work characteristic of all forms of practice. The opening and closing moves in a game of chess, Abbott argues, often appear methodical and rigorous. However, in between these two moves, he argued, is the game itself in which knowledge, expertise, and experience intermingle as the game progresses. On one hand, we can summarize and teach chess as a collection of formal *rules* and techniques (e.g., how a pawn moves, how the rook moves, ways to minimize safe moves for your opponent, etc.). On the other hand, however, we have to acknowledge that any and all *application* of such rules is situated – contingent to the specific layout of the game at hand. In this way, chess (and professional vision) is rule-*based* but not rule-*bound* – a distinction we return to in the discussion.

These insights are backed in turn by a long line of pragmatist social science dealing with the nature of 'routines' and 'routinizable tasks' in organizational and other contexts. Building on Dewey's [12] foundational work, Cohen [9] argues against the common understanding of routinized tasks as collections of rigid and mundane actions, guided by "mindless" rules and mechanized actions; instead, the performance of a routine is both skilled and unique:

> *"For an established routine, the natural fluctuation of its surrounding environment guarantees that each performance is different, and yet, it is the 'same.' Somehow there is a pattern in the action, sufficient to allow us to say the pattern is recurring, even though there is substantial variety to the action."* [9: 782]

Klemp et al. [22] also draw on Deweyan roots to address these "similar, yet different" applications of routines through the vocabulary of plans, takes, and mis-*takes*. There might be a *plan* (a method, an algorithm, a script), and there might be known mistakes (incompatibility, inefficiency, misfit), but every application of the plan is a take ripe for mis-*takes*. Mis-*takes* occur when professionals are faced with something unexpected during the execution of formal and established routines. Drawing on the example of a Thelonious Monk jazz performance, the authors explore the complex discretionary processes by which a musician deals with mis-*takes*:

> *"When we listen to music, we hear neither plans nor mistakes, but takes in which expectations and difficulties get worked on in the medium of notes, tones and rhythms. Notes live in connection with each other. They make demands on each other, and, if one note sticks out, the logic of their connections demands that they be reset and realigned."* [22: 10]

Mis-*takes*, then, mark elements of "contingency, surprise, and repair [found] in all human activities." [22: 4] Signifying the lived differences between theoretical reality and empirical richness, mis-*takes* necessitate situated, often creative, improvisations on the part of professionals and other social actors.

Like Abbott's description of chess, Klemp et al.'s analysis draws out the situated nature of professional knowledge and practice, even in apparently straightforward and routinized procedures. This point is further elaborated by Feldman & Pentland [14], who show how routines are ostensive (the structural rule-like elements of a routine) as well as performative (the situated and contingent execution of a routine). It is the interplay between the two aspects that allows for the discernable but shifting reality of routinized work and professional practice. Along similar lines, Wylie's study of paleontology laboratories [37] shows how adapting situated routines and practices to deal with new problems-at-hand is considered an integral aspect of learning by doing. "Problem-solving in ways acceptable to a field [...] can be an indicator of skill, knowledge, and membership in that particular field." [37: 43]

However, the situatedness of a practice is not always visible. Ingold [17: 98], using the example of a carpenter sawing planks, describes how to an observer, "it may look as though […a] carpenter is merely reproducing the same gesture, over and over again." Such a description, he reminds us, is incomplete:

> *"For the carpenter, [...] who is obliged to follow the material and respond to its singularities, sawing is a matter of engaging 'in a continuous variation of variables…"* [17: 98]

To improvise on seemingly routine tasks then is to "follow the ways of the world, as they open up, rather than to recover a chain of connections, from an end-point to a starting-point, on a route already travelled." [17: 97]

Such social science insights on professional vision and discretionary practice have translated into important CSCW and HCI research programs. For instance, Suchman and Trigg [32] demonstrate the role and significance of representational devices for ways in which Artificial Intelligence (AI) researchers see and produce professional objects and knowledge. Mentis, Chellali, & Schwaitzberg [27] show how laparoscopic surgeons demonstrate ways of "seeing" the body through imaging techniques to students: "seeing" the body in a medical image is not a given, but a process requiring discussion and interpretation. Mentis & Taylor [28] similarly argue that "work required to see medical images is highly constructed and embodied with the action of manipulating the body." Situating objects or phenomena in representations, they argue, is a situated act: representations don't just *reveal* things, but also *produce* them, turning the "blooming, buzzing confusion" of the world [19] into stable and tractable "objects" amenable to analytic and other forms of action.

Performing analytical and other forms of action on the world, however, requires people to deal directly with empirical contingency. Suchman [33] argues that "plans" are theoretical, often formulaic, representations of human actions and practices. "Situated action," however, requires people to work with continuous variation and uncertainty in the world. Human action, she argues, is a form of iterative problem solving in an attempt to accomplish a task. Creativity often emerges within such situated and discretionary forms of problem solving. As Jackson &

Kang's [18] study of interactive artists shows, dealing with material mess and contingency (in this case, attached to the breakdown of technological systems and objects) may necessitate and drive forms of improvisation and creativity at the *margins* of formal order. Creativity – understood not as an abstract and free-standing act of cognition but rather as a *situated sociomaterial accomplishment* – emerges through the interplay between routines and applications, between plans, takes, and mis-*takes*, and between empirical mess and theoretical clarity.

Such situated and discretionary acts are no less central to forms of data analysis and algorithmic knowledge studied and practiced by CSCW and HCI researchers. Clarke [11], for instance, analyzes the human collaborative work in data analytics that is often overlooked in the face of the growing "popularity of automation and statistics." He analyzes the processes used by online advertising professionals to create user models, bringing to light ways in which we can design better software to accommodate the mundane, assumptive, and interpretive deliberation work that goes into producing such "social-culturally constituted" models. Pine & Liboiron [30] study the use of public health data, showing how data collection practices are actually social in nature. One does not simply collect "raw data." Data collection practices are shaped by values and judgments about "what is counted and what is not, what is considered the best unit of measurement, and how different things are grouped together and 'made' into a measurable entity." [30: 3147] Along similar lines, Vertesi & Dourish [36] show how the use and sharing of data in scientific collaboration depends on the contexts of production and acquisition from which such data arise. Taylor et al. [35] show how data materializes differently in different places by and for different actors. Indeed, it is precisely the erasure of these kinds of work that produces the troubling effects of neutrality, "opacity", and self-efficacy that all too often clouds public understanding of "big data," and makes algorithms appear 'magical' in learning, but also 'real-world' environments [8].

These bodies of CSCW and HCI research call attention to aspects of formalism, contingency, and discretion at the heart of algorithmic knowledge and data analytic practices. An algorithm is a collection of formal rules – indeed, a routinizable plan of action – that organizes data in predictable and actionable ways. Yet each dataset poses unique challenges (and opportunities) for the data analyst, necessitating ways to accommodate the variations in the seemingly routine acts of "applying" algorithms. To learn data vision then is to learn to see similarities as well as differences in the ways in which data, algorithms, and worlds are put together. To *see* with data is to see the unknown, the different, and the singular within the space of the mundane and predictable. Advancing an understanding of data analytics as a rule-*based* (as opposed to a rule-*bound*) practice, this paper argues that data vision is not merely a collection of formal and mechanical rules, but a situated and discretionary process requiring data analysts to continuously straddle the competing demands of formal abstraction and empirical contingency.

**METHODS AND FINDINGS**

The arguments that follow build on ethnographic fieldwork conducted at a major U.S. East Coast University. We conducted a four month long participant-observation study in a graduate level machine-learning class taught at the university in fall 2014. One of the authors was enrolled as a student in one section of the course with ~80 students. We also conducted a participant-observation study of a series of three digital humanities workshops organized at the same university during spring 2015. The workshops' purpose was to expose students to computational techniques for text analyses. Each workshop lasted two hours, and the number of participants in each workshop ranged from nine to thirteen.

**CASE 1: MACHINE-LEARNING CLASSROOM**

Our first case follows an instance of data analysis and learning revealed during a machine-learning class. At the point we pick up the story, the instructor is about to introduce a type of algorithm that classifies things into groups (called clusters) such that things within a cluster are sufficiently similar to each other, and things across clusters are sufficiently different from each other.

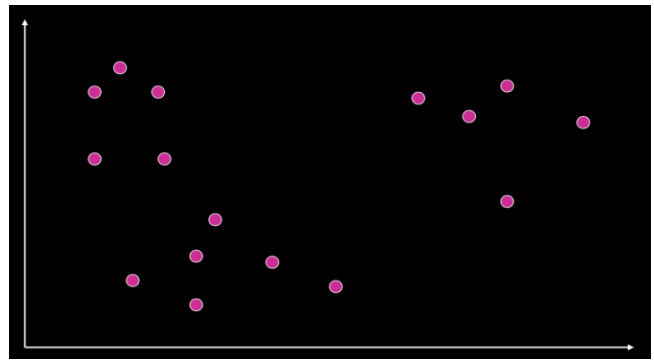

**Figure 1. Class exercise to introduce the notion of clusters.**

The instructor starts by showing an image to the students (figure 1) and inquiring: *how many clusters do you see?* Most students give the same answer: "three clusters."

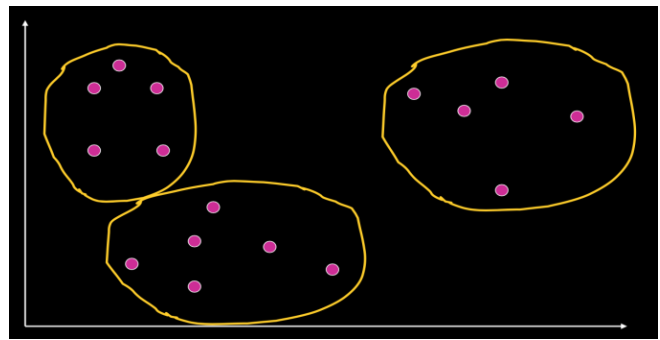

**Figure 2. The three clusters that the students initially saw.**

Having anticipated this response, the instructor shows another image with three clearly marked clusters (figure 2). The instructor informs the students that the number of clusters present in the image is actually unclear:

> *How many clusters? I don't know. I haven't even told you what the similarity measure is* [i.e., how do you even know which two dots are similar to each other in this graph.] *But, you all somehow assumed Euclidean Distance* [i.e., the closer two dots are, the more similar they are.]

He now shows other types of clusters that could have been "seen" (figure 3). As is clear from these images, there could have been two or three clusters. Moreover, there could have been different kinds of two clusters (figure 3a/3b) and different kinds of three clusters (figure 3c/3d). After the students have had a chance to digest this lesson, the instructor goes on to introduce the concept of a clustering algorithm:

> *A clustering algorithm does partitioning. Closer points are similar, and further away points are dissimilar. We haven't yet defined what we mean exactly by similarity, but it's intuitive, right?*

Having made this point, the instructor moves on to a more specific algorithm. The instructor explains that this algorithm works on a simple principle: *the similarity of two clusters is equal to the similarity of the most similar members of the two clusters*.

Having made this point, the instructor moves on to a more specific algorithm. The instructor explains that this algorithm works on a simple principle: *the similarity of two clusters is equal to the similarity of the most similar members of the two clusters*. The idea is to take a cluster (say, X), find the cluster that is most similar to it (say, Y), and then merge X and Y to make a new cluster. It is important to note that knowing the premise on which this algorithm functions is different from knowing how to apply it to data. How do we find a cluster most similar to a given cluster? What does it mean when we say "most similar members of the two clusters"? Such questions, as we will see, are key to this algorithm's application.

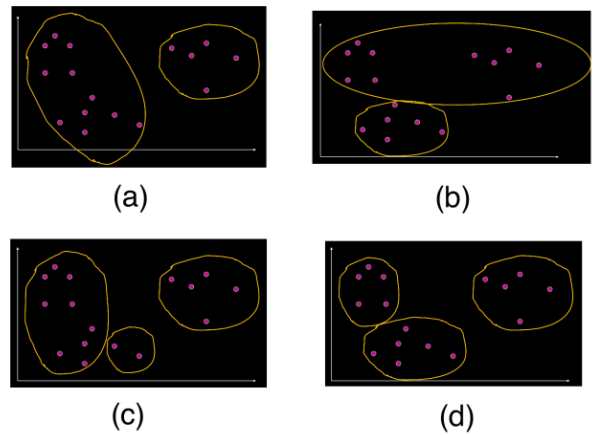

**Figure 3. Different kinds of clusters that could have been seen.**

The instructor now demonstrates the application of this algorithm by drawing a 2-dimensional graph marked with eight dots (figure 4a). The closer the two dots are, he explains, the more similar they are for the purpose of this algorithm. At the start (figure 4a), there are no clusters but only a set of eight dots. The instructor tells the students that each dot will be treated initially as a cluster. He then starts to apply the algorithm beginning with dot-1. On visual inspection, the instructor and students infer that dot-1 is closer to dot-2, dot-3, and dot-4, than it is to the other dots. The instructor and the students then look again, and determine that of the three remaining points, dot-2 is the one closest to dot-1. Thus, based on the chosen similarity metric of physical distance, dot-1 and dot-2 are merged to form cluster-A (figure 4b).

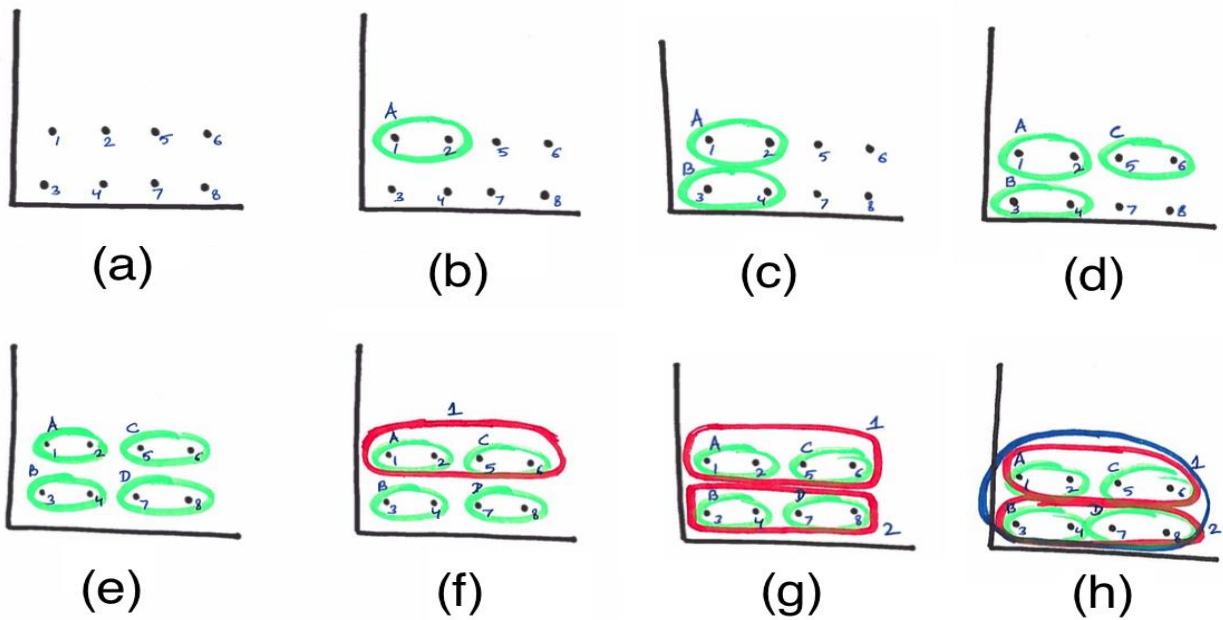

**Figure 4. In-class exercise to learn a particular clustering algorithm.**

The instructor now moves on to dot-3. Following the same logic, the instructor and students infer that dot-3 is closer to cluster-A and dot-4 than it is to the other dots. The instructor reminds the students that for this algorithm, two clusters are compared based on their most similar members (i.e. two dots – one in each cluster – that are closest to each other). Thus, comparing dot-3 and cluster-A, he says, means comparing dot-3 and dot-1 (as dot-1 is the dot in cluster-A that is closest to dot-3). Looking at dot-3, dot-1, and dot-4, the instructor and students infer that dot-4 is the one closest to dot-3; dot-3 and dot-4 are then merged to form cluster-B (figure 4c). In the next two steps, the instructor and students go on to dot-3 and dot-4, forming cluster-C (figure 4d) and cluster-D (figure 4e) respectively.

At this point, eight dots have been lost, and four clusters (with two dots each) gained (figure 4e). After reminding the students that comparing two clusters requires finding two dots – one in each cluster – that are closest to each other, the instructor moves on to cluster-A. A few students point out that the similarity between cluster-A and cluster-B is equivalent to the similarity between dot-1 and dot-3. Other students argue that it is equivalent to the distance between dot-2 and dot-4, as the distances between them look the same. The instructor agrees with the students, and informs them that these distances represent the similarity between cluster-A and cluster-B. The students go on to perform the same analysis to compare cluster-A, -C, and –D.

With regard to cluster-A, the comparison is now down to three sets of distances: between a) dot-2 and dot-4, b) dot-2 and dot-5, and c) dot-2 and dot-7. On visual inspection, the students observe that dot-2 is closest to dot-5. Cluster-A and cluster-C are therefore merged to form cluster-1 (figure 4f). A similar operation merges cluster-B and cluster-D to form cluster-2 (figure 4g). In the last step, cluster-1 and -2 are merged to form a single cluster containing all eight dots (figure 4h). With this, the instructor tells the students, they have reached the end of the exercise, having successfully "applied" the clustering algorithm.

There are three striking features about the in-class exercises described in this section. The first is the step-by-step mechanical nature of the instructor's demonstration of the algorithm. Explicit in the algorithm's demonstration is a collection of formal rules specifying how to treat individual dots, how to compare two dots, how to compare a dot and a cluster, etc. Aspects of data vision, as we see in this case, are built sequentially with students learning an algorithm's application as a set of mechanical and routine steps through which data – represented as dots – are manipulated, enabling the formation of similarity clusters.

A second and related feature is the abstract nature of the represented and analyzed data. These exercises do not have a specific "real-world" context supplementing them. The students were never told, and they never inquired, what the dots and the graph represented. The dots were presented and analyzed simply as label-less dots on a nameless graph, generic representations of any and all kinds of data that this algorithm can work on.

A third and final point concerns the reliance on visuals to demonstrate the operation of the algorithm. We see how visual forms such as dots, circles, and graphs helped students learn to "see" data in ways amenable to formal representation and organization. This allows the students to learn to manipulate the world as a set of data points arrayed in 2-dimensional space. The algorithm, it appears, "works" as long as data is in the form of dots in n-dimensions.

While seeing and organizing the world through mechanical rules and abstract representations is key to data vision, students also need to learn to see the application of an abstract, generic method as a situated and discretionary activity. An instance of this appears in the case below.

## CASE 2: DIGITAL HUMANITIES WORKSHOPS

Our second case follows the construction of data vision as revealed during a series of digital humanities workshops. Digital humanities, broadly put, is a research area in which humanists and information scientists use computational as well as interpretive methods to analyze data in domains such as history and literature. The vignette that follows describes how workshop conveners and students decide what dataset to work on and what happens when they begin to analyze the chosen dataset.

It hasn't been straightforward for the workshop conveners to decide what texts (i.e., data) the students should work on as a group not only because students have different research interests but also because not all texts are digitally available. In the first workshop session, there is a long discussion on how to get digitized version of texts (e.g., from Project Gutenberg, HathiTrust, etc.), what format to use (e.g., XML, HTML, or plain-text files), how to work with specific elements of a file (e.g., headers, tags, etc.), and how to clean the files (e.g., fixing formatting issues, removing stop-words, etc.). The students can, of course, simply download a novel, and start reading it right away, but the point of the discussion is to find ways in which the students can make algorithms do the work of "reading."

While describing ways to convert files from one format to another, something catches the convener's eyes as he shows the students an online novel's source code. There is a vertical bar (|) in certain words such as 'over|whelming' and 'dis|tance.' At first, students suspect the digitized version has not been properly proofread. However, after noticing

more and more words with the vertical bar symbol, the convener returns to the non-source-code version of the novel to discover that these are actually words that cut across lines with a hyphen (-). The computer has been joining the two parts of these words with a vertical bar. At this point, a student asks about ways in which she can recognize such errors, separating "good" from "bad" data. A discussion ensues about ready-to-use scripts and packages. Several students observe that manual reading can help spot such errors, but the whole point of using algorithms is to allow work with much more text than can be read and checked in this way. The discussion ends with no clear answers in sight.

A second question concerns the dataset to be used for purposes of the common class exercises. This decision is reached only by the end of the second session: *English Gothic novels.* This choice is arrived at on the basis of convenience rather than common interest – only one student has a research interest in Gothic literature. But a complete set of English Gothic novels in digital form is perceived to be easier to obtain than other candidates suggested by the group. "The allure of the available," as the convener remarks, "is a powerful thing." But this raises another issue: *what actually qualifies as a Gothic novel?* Something with the word Gothic in the title? One tagged as Gothic by the library? Or one acknowledged as Gothic by the wider literary community? After some discussion, the conveners and students agree to ask one of the library's digital curators to select a set of Gothic novels, and at the start of the third workshop session students are presented with plain-text files of 131 English Gothic novels.

While discussing ways in which this dataset can be used, a student inquires whether it is possible to create a separate file for each novel containing only direct quotes from characters in the novel. The workshop convener and students decide to try this out for themselves and immediately encounter a question: *how can an algorithm know what is and isn't a character quote?* After some discussion, the students decide to write a script that parses the text, inserting a section break each time a quotation mark is encountered. They surmise that this procedure will thereby capture all quotes as the text falling between sequential pairs of quotes. The total of such pairs will also indicate the number of quotes in each novel. Based on this understanding, the students create the below algorithm (in Python) to perform this work:

```
import sys
text = ""

for line in open(sys.argv[1]):
    text += line.rstrip() + " "

quote_segments = text.split("\"")
is_quote = False

for segment in quote_segments:
    print "{0}\t{1}\{2}\n".format("Q" if is_quote else "N", len(segment), segment)
    ## every other segment is a quote
    is_quote = not is_quote
```

When tested against one of the novels in the set however the results are surprising: the script has produced just one section break. Most students feel that this result is "wrong." "Oh wow! That's it?" "I think it didn't even go through the file." "Just one quotation mark?" To see what went wrong, students scroll through the chosen novel, glancing through the first twenty paragraphs or so. Upon inspection, they conclude that there is nothing wrong with their script. It is just that this particular novel actually does not have any quotes in it. (The single quotation mark that the script encountered was the result of an optical character recognition error.) This leads to a discussion of differences in writing styles between authors. A couple of students mention how some authors don't use quotation marks, but instead a series of hyphens (-) to mark the beginning and end of character quotes. This raises a new problem. Is it safe to use quotation marks as proxies for character quotes, or should the script also look for hyphens? Are there still other variations that students will need to account for?

Out of curiosity, the students randomly open a few files to manually search for hyphens. Some authors are indeed using them in place of quotation marks:

*------Except dimity, ------ replied my father.*

Others, however, are using them to mark incomplete sentences:

*But 'tis impossible, ----*

In some cases, hyphens have resulted because em-dashes (—) or en-dashes (–) were converted to hyphens by the optical character recognition system:

*Postscript--I did not tell you that Blandly…*

It is now clear to the students that if hyphens sometimes mark speech, they are less robust than quotation marks as proxies for character quotes. They decide to use only quotation marks for the remainder of the exercise to keep things "relatively simple."

It is now time to choose another novel to test the script. This time, the choice is not so random, as students want a novel that has many character quotes as a "good" sample or test case. The script is changed such that it now parses the text of all the novels, returning a list of novels along with the number of sections produced in each novel. These range from 0 to ~600. Since there are no pre-defined expectations for number of quotes in a novel, there is no way to just look at these numbers and know if they are accurate. However, some students still feel that something has gone "wrong." They argue that because every quote needs two quotation marks, the total number of "correct" quotation marks in a novel should be an even number. By the same logic, the number of sections produced on this basis should also be even. But the result returned shows odd numbers for almost half the novels. Students open some of these "wrong" novels to manually search for quotation marks. After trying this out on five different novels, they are puzzled. The novels do have an even number of quotation marks in them. *Why then is the script returning odd numbers?*

It does not take long to identify the problem. The students are right in pointing out that the number of quotation marks in a novel should be even. However, they have misconstrued how the script creates sections in a novel. A student explains this by reference to one of the novel's in the set: Ann Radcliffe's *The Mysteries of Udolpho.* In the passage below, the python script will go through the text inserting four section breaks:

> *She discovered in her early years a taste for works of genius; and it was St. Aubert's principle, as well as his inclination, to promote every innocent means of happiness. <>"A well-informed mind, <>" he would say, <>"is the best security against the contagion of folly and of vice."<> The vacant mind is ever on the watch for relief, and ready to plunge into error, to escape from the languor of idleness.*

This example shows the students that they had been confusing *sections* with *section-breaks*. Although the script creates four section-breaks in the novel, the number of sections created by the script is actually five. The students realize that the number of sections will thus be one more than the count of quotation marks. Since these will always be even, the number of sections created by the script must always be odd.

The problem has now reversed itself. Whereas earlier the participants believed that an odd number of sections was "wrong", they now agree that having an odd number of sections is actually "right". Why then, they puzzle, do some novels have an even number of sections? The participants manually check out a few "even" novels to search for quotation marks. They discover another set of optical character recognition errors, formatting issues, and variance in authors' writing styles that is producing the "wrong" or unexpected result. At the conclusion of the workshop session shortly thereafter, the students still do not have a script that can reliably extract all character quotes in an automated way.

There are many ways to explain what has happened here. One is to say that the novels were not in the "right" format – they had formatting issues, exhibited style inconsistencies, and contained typographical errors. This, however, is true for most, if not all, kinds of data that analysts have to deal with on a daily basis. Clean, complete, and consistent datasets – as every data analyst knows – are a theoretical fantasy. Outside of theory, data is often inconsistent and incomplete. The requirement of prim and proper datasets, we argue, does not do justice either to the reality of the data world or to the explanation of this workshop exercise.

Another explanation is that the students simply lacked skill and experience, and were making what some would call "rookie mistakes". After all, these students were here to learn these methods, and were not expected to know them beforehand. However, the ability to identify and avoid "rookie mistakes" is in itself an important artifact of the training and professionalization of would-be professionals. In large part, what makes a rookie a rookie is his/her inability to recognize and avoid these kinds of errors. As sites for learning and training, classrooms and workshops thus provide avenues for seeing how would-be professionals learn to "see" and avoid "rookie mistakes." Similar if less stark examples of such mistakes appeared in the machine-learning class (using part of training data as a test case, confusing correlation for causation, etc.).

Our workshop case brings together prior knowledge, human decisions, and empirical contingency. The choice of the dataset is not a given, but a compromise between thematic alignment and practical accessibility. Moreover, as seen in the case of vertical bars, hyphens, and quotation marks, data is often idiosyncratic in its own ways, necessitating situated and discretionary forms of pre-processing. Even clearly articulated computational routines (e.g., search for quotation marks, label text between marks as a section, count sections, put sections in a separate file) often require a host of situated decisions (e.g., what novels to look at, what stylistic elements to account for, how to alleviate formatting errors, how to infer and manage empirical contingency, etc.). In all these ways, algorithmically identifying and extracting character quotes is a situated activity that requires practitioners to find their way around specificities of the data at hand.

**DISCUSSION**

The cases above provide important insight into the practice and professionalization of would-be data analysts. In case one, we saw how machine learning students learn to see data in forms amenable to algorithmic manipulation, and an algorithm's application as a collection of formal rule-like steps. The rules to be followed appear methodical, rigorous, and mechanical, and the algorithm is demonstrated using an abstract representational form: label-less dots on a name-less graph. Whether it is discerning the similarity between two dots or knowing ways to compare and merge clusters of dots, students learn to work with and organize the world through a fixed set of rules. Such a demonstration privileges an abstract understanding of data analytics, allowing students to learn to manipulate the world in predictable and actionable ways. This, we argue, is a great source of algorithmic strength: if the hallmark of real-world empirics is its richness and unpredictability, the hallmark of data analysis is its ability to organize and engage the world via abstract categorization and computationally actionable manipulation.

In case two, by contrast, we saw how processes of learning and practicing data analysis are also situated, reflexive, and discretionary, in ways that abstract representations and mechanical demonstrations significantly understate. Multiple decisions were required to effectively combine the script with the given dataset ranging from identifying how to isolate character quotes, discerning ways in which quotes appear in data, to figuring out how to test the script. Unique datasets necessitate different fixes and workarounds, requiring a constant adjustment between prior knowledge, empirical contingencies, and formal methodologies. Making prior knowledge and abstract methods work with data is indeed hard work. Data may be hard to find, unavailable, or incomplete. Under such circumstances, practitioners have to make do with what they can get, in ways that go against the abstracted application story usually shared in data analytic research papers and presentations.

Recognizing the incomplete nature of the abstracted data story helps situate an algorithm's application as a site not only for abstract categorization and formal manipulation but also for *discretion* and *creativity*. Learning to apply an algorithm, as we saw, involves a series of situated decisions to iteratively, often creatively, adapt prior knowledge, data analytic routines, and empirical data to each other. Elements of creativity manifest themselves as professional acts of discretion in the use of abstract, seemingly mechanical methods. While certain datasets may share similarities that support mechanical applications of rules across contexts, mastery of operations in their mechanical form constitutes only one part of the professionalization of data analysts. Each dataset is incomplete and inconsistent in its own way, requiring situated strategies, workarounds, and fixes to make it ready and usable for data analysis. Data analysts are much like Suchman's [33] problem solvers, Klemp et al.'s [22] musicians, and Ingold's [17] carpenters: constantly negotiating with and working around established routines in the face of emergent empirical diversity.

Viewing data analysis as an ongoing negotiation between rules and empirics helps mark a clear distinction between two ways of describing the professionalization and practice of data analytics that are relevant for CSCW and HCI researchers. One of these approaches data analytics as a rule-*bound* practice, in which data is organized and analyzed through the application of abstract and mechanical methods. Casting data analytics as a rule-bound practice helps make visible specific aspects of data analytic learning and practice. First, it allows data analysts to better understand the abstract nature of data analytic theories, facilitating novel ways of computationally organizing and manipulating the world. Second, it enables researchers to focus on constraints and limits of algorithmic analyses, providing a detailed look at some of the critical assumptions underlying data analyses. Finally, it allows students to learn not only how to work with basic, yet foundational, data analytic ideas, but also how to organize and manipulate the world in predictable and actionable ways. However, the same properties that make these aspects visible, tend to render *in*-visible the empirical challenges confronting efforts to make algorithms work with data, making it difficult to account for the situated, often creative, decisions made by data analysts to conform empirical contingency to effective (and often innovative) abstraction. What's left is a stripped down notion of data analytics – analytics as rules and tools – that only tells half the data analytic story, understating the breadth and depth of human work required to make data speak to algorithms. Significantly underappreciating the craftsmanship of data analysts, the rule-bound perspective paints a dry picture of data analysis – a process that often comprises of artful and innovative ways to produce novel forms of knowledge.

A more fruitful way to understand data analytics, we argue, is to see it not as rule-bound but rather as rule-*based*: structured but not fully determined by mechanical implementations of formal methods. In a rule-bound understanding, an algorithm's application requires organization and manipulation of the world through abstract constructs and mechanical rules. In a rule-based understanding, however, emergent empirical contingencies and practical issues come to the fore, reminding us that the world requires a large amount of work for it to conform to high-level data analytic learning, expectations, and analyses. Following Feldman & Pentland's [14] view of routines, a rule-based understanding of data analysis casts algorithms as ostensive as well as performative objects, highlighting how the performances of algorithms draw on and feed into their ostensive nature, and vice versa.

Seeing data analytics as a rule-based practice focuses our attention on the situated, discretionary, and improvisational nature of data analytics. It helps make salient not only the partial and contingent nature of the data world (i.e., data is often incomplete and inconsistent), but also the role of human decisions in aligning the world with formal assumptions and abstract representations of order as stipulated under abstract algorithmic methods and theories. Data analysis is a craft, and like every other form of craft it is never fully *bound* by rules, but only *based* on them. A rule-based understanding of data analysis acknowledges and celebrates the lived differences between theoretical reality, empirical richness, and situated improvisations on the part of data analysts.

It is in and through these lived differences that data analysts gain data vision. As with Dewey's [12], Cohen's [9], and Feldman & Pentland's [14] descriptions of routines and routinized tasks, we see in data vision the always-ongoing negotiation between abstract algorithmic "routines" and the situated and reflexive "applications" of such "routines." Data vision is much like an array of plans, takes, and mis-*takes* [22] – a constant reminder of the situated and discretionary nature of the professionalization and practice of data analysis.

Such an understanding of data vision can inform data analytic learning, research, collaboration, and practice in three basic ways. First, it helps focus attention on the role of human work in the professionalization and practice of data analytics; while models, algorithms, and statistics clearly matter, focusing on situated and discretionary judgment helps contextualize algorithmic knowledge, facilitating a better understanding of the mechanics, exactness, and limits of such knowledge. Algorithms and data don't produce knowledge by themselves. We produce knowledge *with* and *through* them. The notion of data vision puts humans back in the algorithm.

Second, data vision can help us better attend to the ways in which algorithmic results are documented, presented, and written up. Although algorithmic and statistical choices constitute a significant part of data analytic publications, also providing an explicit description of key decisions that data analysts take can not only help communicate a nuanced understanding of technical choices and algorithmic results, but also enable students as well as practitioners to think through aspects of their work that though may seem "non-technical," greatly impact their knowledge claims. This helps to not only reduce the "opacity" [8] of data analytic practices, but also better teach and communicate, what some call, the "black art" or "folk knowledge" [13] of data analysis, contributing to the development of a complete and "reflective practitioner" [31].

Third, better understanding of data vision can help inform both professional training and community conversations around data analysis. In data analytics, and in many other forms of research (including our own!), we often present research setup, process, and results in a dry and straightforward manner. *We had a question, we collected this data, we did this analysis, and here is the answer.* Open and effective conversations about the messy and contingent aspects of research work – data analytic or otherwise – tend to escape the formal descriptions of methods sections and grant applications, reserved instead for water cooler and hallway conversations by which workarounds, 'tricks of the trade', and 'good enough' solutions are shared. The result is an excessively "neat" picture that fails to communicate the real practices and contingencies by which data analytic work proceeds. This becomes even more difficult outside the classroom. In industry, research centers, and other contexts of algorithmic knowledge production, data analysts often work with huge volumes of data in multiple teams, simultaneously interfacing with a host of other actors such as off-site developers, marketers, managers, and clients. Where the results of data analytics meet other kinds of public choices and decisions (think contemporary debates over online tracking and surveillance, or the charismatic power of New York Times infographics) these complications – and their importance – only multiply. Data analytic results often travel far beyond their immediate contexts of production, taking on forms of certainty and objectivity (even magic!) that may or may not be warranted, in light of the real-world conditions and operations from which they spring. Here as in other worlds of expert knowledge, "distance lends enchantment" [10].

More broadly, an understanding of data vision helps support the diverse forms of oft-invisible collaborative data analytic work. Data analysis not only warrants algorithmic techniques and computational forms, but also comprises answers to crucial questions such as what is the relation between data and question, what can actually be answered through data, what are some of the underlying assumptions concerning data, methods, etc. By bringing such questions – and, indeed, other forms of human work – to the fore, data vision directs our attention to forms of situated discretionary work enabling and facilitating data analysis. Data are never "raw" [15], and a large amount of work goes into making data speak for themselves. The notion of data vision can help us to identify and build acknowledgment and support mechanisms for sharing such folk knowledge that, though immensely useful, is often lost. Data vision is not merely about perceiving the world, but a highly consequential way of seeing that turns perception into action. Data often speak specific forms of knowledge to power. Like all forms of explanation, data analysis has its own set of biases [3, 15], assumptions [4, 7, 34], and consequences [4, 5, 6]. Understanding data vision allows us to better delineate and communicate the strengths as well as the limitation of such collaborative knowledge – indeed, of *seeing* the world with and through data.

**CONCLUSION**

Given our growing use of and reliance on algorithmic data analysis, an understanding of data vision is now integral to contemporary knowledge production practices, in CSCW and indeed many other fields. In this paper we presented two distinct, yet complementary, ways of learning and practicing data analysis. We argued in favor of a rule-based, as opposed to a rule-bound*,* understanding of data analytics to introduce the concept of data vision – a notion that we find integral, if not foundational, to the professionalization and practice of data analysts. We described how a better understanding of data vision allows us to better grasp and value the intimate connection between methodological abstraction, empirical contingency, and situated discretion in data analytic practice. Shedding light on the diverse forms of data analytic work, data vision produces a more open and accountable understanding of algorithmic work in data analytic learning and practice.

Studying learning environments helps showcase basic, yet formative, aspects in the training and professionalization of data analysts. In this paper, using empirical examples from classrooms and workshops, we have described not only a rule-based view of data analysis, but also the outline of the notion and practice of data vision. Studying learning environments, however, has its limitations. Classrooms are but one step in the professionalization of data analysts. Data analysis, like all practices, is a constant learning endeavor. To better understand data analytic practice, we then need to also study other contexts of algorithmic knowledge production such as those in industry, research centers, startups, and even hackathons. Acting as avenues for future research, diverse contexts of data analyses provide opportunities to further and strengthen our understanding of data vision. In different contexts, data analysis is shaped by a diverse set of professional expectations and organizational imperatives, reminding us that the practice of data analysis remains a deeply social and collaborative accomplishment. This paper has suggested early steps in defining and understanding data vision. Future work will seek to extend and deepen this holistic approach.

**ACKNOWLEDGMENT**

Support for this research was provided by U.S. National Science Foundation Grant #1526155, and the Intel Science and Technology Center (ISTC) for Social Computing.